\documentclass[aps,prl,nofootinbib,superscriptaddress,preprintnumbers,amsmath,amssymb,latexsym,array,enumerate,letterpaper,twocolumn]{revtex4}
\usepackage{epsfig}
\usepackage[colorlinks=true,urlcolor=RoyalBlue,linkcolor=Blue,citecolor=magenta]{hyperref}
\usepackage{breakurl}
\usepackage{lineno}
\usepackage{bm,color}
\usepackage[dvipsnames]{xcolor}
\usepackage{slashed} 
\usepackage{graphicx}
\usepackage{soul} 
\usepackage{multirow}
\usepackage{comment}
\usepackage{epstopdf} 
\usepackage{mathtools}
\usepackage{appendix}
\usepackage{url}
\usepackage{doi}
\usepackage[english]{babel}
\usepackage{ulem}

\makeatletter
\def\simgt{\mathrel{\lower2.5pt\vbox{\lineskip=0pt\baselineskip=0pt
           \hbox{$>$}\hbox{$\sim$}}}}
\def\simlt{\mathrel{\lower2.5pt\vbox{\lineskip=0pt\baselineskip=0pt
           \hbox{$<$}\hbox{$\sim$}}}}
\makeatother

\newcommand{\be}{\begin{equation}}
\newcommand{\ee}{\end{equation}}
\newcommand{\bea}{\begin{eqnarray}}
\newcommand{\eea}{\end{eqnarray}}
\newcommand{\beq}{\begin{eqnarray}}
\newcommand{\eeq}{\end{eqnarray}}
\newcommand{\Fig}[1]{Fig.~\ref{#1}}
\newcommand{\Eq}[1]{Eq.~(\ref{#1})}

\newcommand{\mPl}{M_{\rm P}}
\newcommand{\mevp}{M_\textrm{evp}}

\newcommand{\dd}[1]{\frac{\partial}{\partial #1}}

\newcommand{\BH}{\textrm{BH}}

\def\lsim{\mathrel{\rlap{\lower4pt\hbox{\hskip1pt$\sim$}}
     \raise1pt\hbox{$<$}}}         
\def\gsim{\mathrel{\rlap{\lower4pt\hbox{\hskip1pt$\sim$}}
     \raise1pt\hbox{$>$}}}         

\begin{document}

\widetext
\preprint{UMD-PP-026-04, MI-HET-885}

\title{Probing New Degrees of Freedom with the Universal Tail\\of Primordial Black Hole Mass Functions}

\author{Kaustubh~Agashe}
\email{kagashe@umd.edu}
\affiliation{Maryland Center for Fundamental Physics, Department of Physics, University of Maryland, College Park, MD 20742, USA}

\author{Jae~Hyeok~Chang} 
\email{jhchang056@snu.ac.kr}
\affiliation{Center for Theoretical Physics, Department of Physics and Astronomy,
Seoul National University, Seoul 08826, Korea}

\author{Steven~J.~Clark} 
\email{sclark@hood.edu}
\affiliation{Department of Chemistry and Physics, Hood College, Frederick, MD 21701, USA}

\author{Bhaskar~Dutta} 
\email{dutta@tamu.edu}
\affiliation{Mitchell Institute for Fundamental Physics and Astronomy, Department of Physics and Astronomy, Texas A\&M University, College Station, TX 77845, USA}

\author{Yuhsin~Tsai} 
\email{ytsai3@nd.edu}
\affiliation{Department of Physics and Astronomy, University of Notre Dame, IN 46556, USA}

\author{Tao~Xu} 
\email{tao.xu@ust.hk}
\affiliation{Jockey Club Institute for Advanced Study, Hong Kong University of Science and Technology, Clear Water Bay, Kowloon, Hong Kong}
\affiliation{Department of Physics and Astronomy, University of Oklahoma, Norman, OK 73019, USA}

\begin{abstract}
The primordial black hole (PBH) mass function today develops a low-mass evaporation tail whose shape is universal with respect to the initial PBH mass distribution. This universality is fixed by the continuity equation and the Hawking mass-loss rate, but the tail is distorted if additional particle degrees of freedom participate in Hawking evaporation. Since this tail controls high-energy PBH photon emission, such distortions leave characteristic features in the gamma-ray spectrum. We show that these features provide a robust probe of new degrees of freedom, even for subdominant PBHs, within reach of future experiments and independent of visible-sector couplings or relic abundance.
\end{abstract}

\maketitle

\section{Introduction}
Primordial black holes (PBHs) offer a unique window into both early-Universe cosmology and fundamental particle physics. PBHs can form in the early Universe~\cite{Carr:1974nx} and span a wide range of masses~\cite{Carr:2026hot}, including masses small enough to have Hawking temperatures far above those of astrophysical black holes. Since Hawking radiation~\cite{Hawking:1975vcx} is a gravitational process, PBHs emit all particle species kinematically accessible at their Hawking temperature, regardless of their non-gravitational interactions. PBHs therefore act as democratic emitters and provide a powerful and model-independent probe of physics beyond the Standard Model (SM).

A particularly important mass scale is $\mevp \sim 10^{15}$~g, defined as the PBH mass whose lifetime equals the age of the Universe. PBHs with present-day masses below $\mevp$ are actively evaporating today and can contribute to observable radiation backgrounds, most notably gamma rays. The evolution of the PBH mass function is governed by a continuity equation in mass space, with the mass-loss rate set by Hawking radiation. This equation implies that the present-day PBH mass function below $\mevp$ develops a universal evaporation tail, whose shape is independent of the mechanism responsible for the gravitational collapse into PBHs. The formation history enters only through an overall normalization set by the initial PBH abundance at $\mevp$, since the PBHs populating the tail today originate from an extremely narrow mass band around $\mevp$. 

Here, ``universality'' means that the shape of the low-mass PBH tail is insensitive to the initial PBH mass function. Within the SM, the evaporation rate and hence the shape of the mass tail are uniquely determined. However, additional particle degrees of freedom (dof), if present, accelerate the mass loss and thereby modify the shape of the tail away from its SM form. This effect depends only on the number of available dof and their mass thresholds, and is insensitive both to their interactions with the visible sector and to the reheating history of the Universe. The evaporation tail leaves a direct imprint on the gamma-ray spectrum, so deviations from its SM shape allow the presence of new dof to be inferred observationally.

In this work, we demonstrate for the first time that the spectral structure of the seemingly universal PBH evaporation tail constitutes a powerful probe of new dof, even when PBHs make up only a subdominant fraction of dark matter (DM). Although the present-day PBH evaporation tail has appeared in previous analyses~\cite{Mosbech:2022lfg,Klipfel:2025bvh, He:2026neq}, these works have considered only SM particle content. Refs.~\cite{Baker:2021btk,Baker:2022rkn,Baker:2025ffi} study how additional particle species accelerate black hole evaporation, but focus on signals from individual black holes rather than the mass tail. We analyze the gamma-ray spectra from the mass tail for scenarios with large additional numbers of dof at different mass scales, and find that the required telescope effective area and observation time for probing $\mathcal{O}(10\text{--}10^3)$ new dof below the $10$~GeV scale are comparable to those of the ongoing Cherenkov Telescope Array (CTA) observatory~\cite{Lopez-Oramas:2025vld} and proposed missions such as the Advanced Particle-astrophysics Telescope (APT)~\cite{Buckley2019Advanced,APT:2021lhj,Alnussirat:2021tlo} and the Very Large Area gamma-ray Space Telescope (VLAST)~\cite{Zhang:2024jha}. PBH evaporation therefore serves as a robust calorimeter for new particle dof, enabling tests of well-motivated beyond-the-Standard-Model (BSM) frameworks containing large numbers of light dof, such as axion-like particles (ALPs) in axiverse models~\cite{Arvanitaki:2009fg,Dessert:2025yvk} or SM-like particles in Mirror-Twin-Higgs~\cite{Chacko:2005pe,Chacko:2005un}, $N$naturalness models~\cite{Arkani-Hamed:2016rle}, and mirror-DM~\cite{Foot:2014mia} models.

\section{PBH Mass Function and Its Tail}
After PBHs are formed in the early Universe at $t=t_i$, they lose mass through Hawking evaporation. Assuming negligible merger and accretion rates, their evolution is governed by the continuity equation for the comoving number density per unit mass, $\hat{n}_{\BH}(t,M) \equiv a^3 \, {\rm d}n_\BH / {\rm d}M$, where $a$ is the scale factor, $n_\BH$ is the physical number density of PBHs, and $M$ is the PBH mass:
\beq\label{eq:nBHPDE}
{\dd t} \hat{n}_\BH(t,M) + {\dd M} \left( \dot{M}(M) \, \hat{n}_\BH(t,M) \right) = 0 \,,
\eeq
where the mass-loss rate is given by
\bea
\dot{M}(M) = - \frac{\mathcal{G} g_{\star H}(T_\BH(M)) \mPl^4}{30720 \pi \, M^2}
= - \varepsilon(M) \frac{\mPl^4}{M^2} \, .
\label{eq:massloss}
\eea
Here, $\mPl$ is the Planck mass, $\mathcal{G}$ is the numerical factor arising from the greybody average over particle species, and $g_{\star H}(T_\BH)$ denotes the effective Hawking degrees of freedom, including the species accessible at the Hawking temperature $T_\BH (M) = 1/(8 \pi G_{\textrm{N}} M)$ with $G_{\mathrm{N}}$ as the gravitational constant. A more detailed description can be found in~\cite{Cheek:2022mmy} from which we take the numerical values of $\varepsilon(M)$. From \Eq{eq:nBHPDE} and \Eq{eq:massloss}, we  obtain the PBH number density per mass today ($t=t_0$),
\begin{eqnarray}
    \hat{n}_{\BH}(t_0,M) = \frac{\varepsilon(M_i)}{\varepsilon(M)} \frac{M^2}{M_i^2} \hat{n}_{\BH}(t_i,M_i)\, ,
\label{eq:density_comp}
\end{eqnarray}
where $M_i(M)$ is the initial mass of a PBH whose present mass is $M$. We provide a detailed derivation of \Eq{eq:density_comp} in the {\it Supplemental Material}. 

Since the evaporation rate $\dot{M} \propto 1/M^2$ accelerates as the PBH mass decreases, a PBH spends only a brief final period at masses well below its initial value. Consequently, the PBHs found in the tail at $M \ll \mevp$ today must have started with initial masses $M_i \approx \mevp$.
Therefore, the tail of the PBH mass function can be obtained by replacing $M_i$ in \Eq{eq:density_comp} with $\mevp$, so that the only remaining mass dependence is $M^2/\varepsilon(M)$.
Only the overall amplitude depends on the early-Universe history, establishing the universality of the tail shape. Apart from the factor $\varepsilon(M)$, which encodes the BSM degrees of freedom, the tail scales as $M^2$. Additional particle species beyond the SM with mass below $T_\BH(M)$ contribute to $g_{\star H}$, modifying $\varepsilon(M)$ and the tail shape, regardless of how these particles couple to the visible sector.

\begin{figure*}[t]
\centering
\includegraphics[width=\columnwidth]{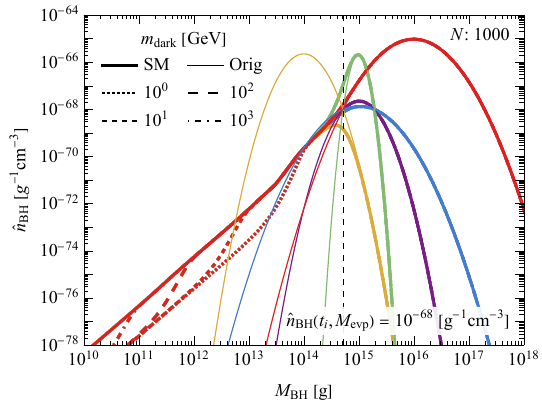}
\includegraphics[width=\columnwidth]{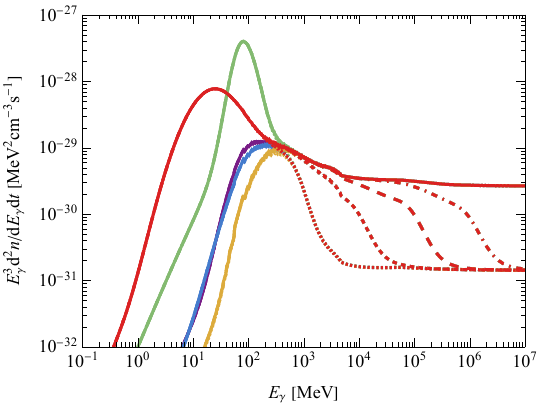}
\caption{\textbf{Left:} Present-day PBH distribution $\hat{n}_\textrm{BH}(t_0,M)$ for various initial PBH distributions. The thin lines show log-normal initial mass spectra $\hat{n}_\textrm{BH}(t_i,M)$ with different $\mu$ and $\sigma$ (see \Eq{eq:lognormaldist}), each in a distinct color. All initial spectra have identical densities at $\mevp$, corresponding to $f_\text{PBH}\approx 2\times 10^{-8}$ for the present-day {\it purple} curve. The solid curves assume only SM particle content while the various dotted and dashed curves include $N=10^3$ additional scalar dof with mass $m_{\rm dark}$ as indicated in the legend. For $M<\mevp$, all solid curves converge to a single universal tail. The enhanced evaporation rate from the extra dof alters the tail relative to the SM prediction, with the onset of the deviation determined by $m_{\rm dark}$. \textbf{Right:} Gamma-ray spectrum weighted by $E_{\gamma}^3$ for the present-day distributions presented in the {\it left} panel, with line styles and colors matching. The gamma-ray spectra at $E_\gamma > 1$~GeV are nearly identical across initial conditions, reflecting the universality of the tail, while additional dof produce a clean BSM signature with an onset set by $m_\mathrm{dark}$. Differences in the initial mass function affect only the sub-GeV region.
}
\label{fig:BHbenchmark}
\end{figure*}

The left panel of \Fig{fig:BHbenchmark} illustrates this mass-tail universality. We model the initial distributions as log-normal,
\beq\label{eq:lognormaldist}
\hat{n}_\textrm{BH}(t_i,M) = \frac{A}{\sqrt{2\pi}\,\sigma M}\exp\left[-\frac{(\sigma^2-\log[M/\mu])^2}{2\sigma^2}\right] \, ,
\eeq
where $\mu$ is the peak mass, $\sigma$ the width, and $A$ the overall normalization. We take several such spectra with identical amplitudes at $\mevp$ but different $\mu$ and $\sigma$ (thin lines of different colors) and evolve them to their present-day distributions (thick lines) using \Eq{eq:density_comp}. For $M < \mevp$, the solid curves, computed with SM particle content, collapse onto a single universal tail, independent of the formation history. The dashed and dotted curves show the effect of $N = 10^3$ additional dof with mass threshold $m_{\rm dark}$. These extra species contribute to Hawking radiation once $T_{\rm BH}(M) \gtrsim m_{\rm dark}$, enhancing $\varepsilon(M)$ in \Eq{eq:massloss} and suppressing the tail relative to the SM prediction. This deviation establishes a direct correspondence between the new particle mass and the observable modification of the tail.

The total PBH energy density $\rho_\textrm{PBH}$ relative to the cosmic DM energy density $\rho_\textrm{DM}$,
\bea
f_\textrm{PBH} = \frac{\rho_\textrm{PBH}}{\rho_\textrm{DM}}
&=& \frac{1}{\rho_\textrm{DM}}\int M  \hat{n}_\BH(t_0, M) \, \textrm{d}M \, ,
\eea
sets the overall amplitude of the present-day mass function. We show below that the BSM signature in the evaporation tail is detectable even for $f_\mathrm{PBH}\ll1$.

\section{Gamma-ray Spectrum from PBH Tail}\label{sec.gammarayPBH}
Hawking radiation is one of the most direct PBH detection channels. Particle emission from a Schwarzschild PBH follows an approximately blackbody spectrum at temperature $T_\mathrm{BH}(M)$, modified by energy- and spin-dependent greybody factors and Boltzmann suppressed for species heavier than $T_\mathrm{BH}$. Since Hawking evaporation depletes angular momentum much faster than mass~\cite{Page:1976ki}, we neglect PBH spin and the associated spectral corrections, which in any case do not affect the universal feature discussed below.

Among accessible particle species, photons, both primary from direct emission and secondary from subsequent processes, enable the most accurate reconstruction of the PBH mass function. Therefore, we focus on the gamma-ray emission from PBHs, as its spectrum directly reflects the present-day PBH mass function. The photon emission rate per unit volume is obtained by integrating the single-PBH spectrum, $\textrm{d}^2N_\gamma/\textrm{d}E_{\gamma}\textrm{d}t$, 
over the PBH mass,
\begin{equation}\label{eq:dndEdt}
    \frac{\textrm{d}^2n_{\gamma}}{\textrm{d}E_{\gamma}\textrm{d}t}=\int \hat{n}_\BH(t_0, M) \frac{\textrm{d}^2N_\gamma}{\textrm{d}E_{\gamma}\textrm{d}t} \,\textrm{d}M \, .
\end{equation}
Although \Eq{eq:dndEdt} uses the cosmic-averaged PBH density, the spectral shape is independent of the overall density and is therefore the same in any DM-rich region.

The PBH mass-function tail exhibits the universal scaling shown in Eq.~(\ref{eq:density_comp}). 
The tail mainly scales as $\hat{n}_{\rm BH}\propto M^2$, which gives rise to a high-energy gamma-ray spectrum $\textrm{d}^2n_{\gamma}/\textrm{d}E_{\gamma}\textrm{d}t \propto E_\gamma^{-3}$ for $E_\gamma \gtrsim 1~{\rm GeV}$, distinct from the $\mathcal{O}({\rm MeV})$ gamma rays emitted by heavier PBHs. The definite spectral slope is a characteristic signature of ongoing PBH evaporation.

Additional dof in the Hawking emission modify $\varepsilon(M)$ and leave a direct imprint on the mass function tail, which in turn affects the observed gamma-ray spectrum. New states are produced once the PBH mass drops low enough that its Hawking temperature exceeds the particle mass, yielding a direct correlation between the new mass scale, the associated PBH mass experiencing excess evaporation, and the photon energy where alteration appears. Even if these states are completely secluded, they still indirectly induce observable distortions in the gamma-ray spectrum, in close analogy to new-physics probes through anomalous cooling channels in astrophysical objects~\cite{Raffelt:1990yz, Raffelt:1996wa}.

In the right panel of \Fig{fig:BHbenchmark}, we show spectral suppression induced by the accelerated evaporation into new dof, as encoded in $\varepsilon(M)$ of \Eq{eq:massloss}. Assuming $N=10^3$ new states with fixed mass $m_{\textrm{dark}}$ from $\textrm{GeV}$ to $\textrm{TeV}$, the gamma-ray flux drops by over an order of magnitude at $E_{\gamma} \gg m_{\textrm{dark}}$, compared to the SM scenario. This signature is particularly clean at $E_{\gamma}\gtrsim 1~\textrm{GeV}$, where the overlap of the colored curves demonstrates dominant dependence on the particle model and insensitivity to the formation mass function, except in extreme cases. Joint flux measurements along the tail at multiple energies thus test this effect and identify new particles at GeV scales and above via deviations from the universal power-law scaling in the SM.

\section{Sensitivity to PBH Tail Modifications}

\begin{figure}[t]
\includegraphics[width=\columnwidth]{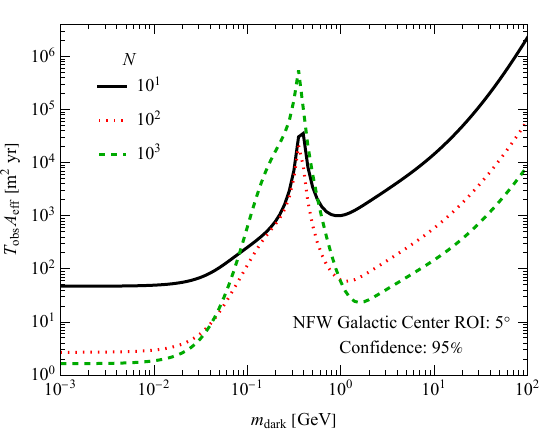}
\caption{An estimate of the minimum observation time ($T_\mathrm{obs}$) times effective area ($A_\mathrm{eff}$) to probe BSM models with additional dof using the tail of a PBH distribution. Individual lines correspond to models with $N$ additional scalar dark dof with mass $m_\mathrm{dark}$. At high masses, sensitivity is lost due to differences occurring deeper in the tail where there is less signal, while at low masses, the limit asymptotes due to BSM particles becoming massless compared with $T_\mathrm{BH}(\mevp)$. For middle masses, the sensitivity decreases due to the spectral drop from additional dof occurring before the observation window, leading to a reduced signal.}
\label{fig:telescope_spec}
\end{figure}

Here we study the detectability of PBH mass tail alterations from BSM dof through modifications to gamma-ray emission, highlighting the discovery potential of the tail region even when the underlying PBH mass distribution is unknown or the new particles do not interact with the SM.

We analyze gamma rays from the Milky Way's Galactic Center (GC), assuming a Navarro–Frenk–White (NFW) DM halo profile~\cite{Navarro:1996gj} with a region of interest (ROI) of $5^\circ$, and compute the expected flux from the emission rate of \Eq{eq:dndEdt} weighted by the observation geometry,
\begin{equation}\label{eq:dPhidE}
    \frac{\textrm{d}\Phi_{\gamma}}{\textrm{d}E_{\gamma}}=f_\mathrm{PBH} \frac{\bar{J}_D}{\rho_\textrm{PBH}} \frac{\Delta \Omega}{4 \pi} \frac{\textrm{d}^2n_{\gamma}}{\textrm{d}E_{\gamma}\textrm{d}t} \,,
\end{equation}
where $\Delta\Omega$ is the solid angle of the ROI and $\bar{J}_D = 1.597\times 10^{26}~\textrm{MeV}~\textrm{cm}^{-2}~\textrm{sr}^{-1}$~\cite{deSalas:2019pee,Coogan:2020tuf} is the angle-averaged $D$-factor of the observation region. Note that the emission rate in \Eq{eq:dndEdt} is evaluated for a reference PBH distribution of total density $\rho_{\rm PBH}$, but since the rate is linear in the PBH number density, the ratio of the two is independent of the chosen $\rho_{\rm PBH}$, while $f_{\rm PBH}$ sets the overall flux normalization. The PBH photon emission spectrum is obtained with BlackHawk~\cite{Arbey:2019mbc, Arbey:2021mbl}. Since the tail's shape is universal and the initial distribution affects only its overall amplitude, our analysis does not depend on the specific PBH distribution except for ensuring consistency with existing bounds.

As a baseline, we adopt the log-normal distribution introduced in \Eq{eq:lognormaldist} with $\sigma=0.5$ and $\mu=10^{15}~\textrm{g}$ (the {\it purple} curve in \Fig{fig:BHbenchmark}), with $A$ set to saturate current experimental bounds for each evolved model.\footnote{Specifically, $A$ is set so that $E_\gamma^2 {\rm d}\Phi_\gamma / {\rm d}E_\gamma$ at 100 MeV is $2.5\times 10^{-5}~{\rm MeV~cm^{-2}~s^{-1}}$, comparable to the Fermi-LAT isotropic diffuse gamma-ray intensity~\cite{Fermi-LAT:2014ryh}.} For SM particle content, this corresponds to $f_\text{PBH} = 2.1 \times 10^{-8}$. 
Although PBHs are discrete objects whose number density decreases toward lower masses, there are at least $\mathcal{O}(10^3)$ PBHs within the Galactic Center observation volume in a mass bin even around $10^{10}$~g, providing sufficient statistics for our analysis.
Therefore, the calculation using the smooth-limit value of $\bar{J}_D$ remains valid across the tail region considered in this work.

To assess the discovery potential of the tail, we evolve the baseline distribution by using \Eq{eq:density_comp} with $\varepsilon(M)$ from a given BSM particle content and compare the resulting photon spectrum with that obtained from the SM content. We bin both the SM and BSM spectra into 10 energy bins per decade\footnote{This binning approximates current experimental energy resolutions~\cite{Caputo:2022xpx,Fermi-LAT:2009ihh}.} and discard all bins below 1 GeV, since the tail's gamma-ray spectrum stabilizes by this energy, as shown in the right panel of \Fig{fig:BHbenchmark}. Because the signal falls as $E_\gamma^{-3}$, we also discard bins above 10 GeV. To compare the two spectra, we rescale the SM spectrum to the BSM's value at the 1 GeV bin so that the comparison reflects only differences in spectral shape rather than overall amplitude in the tail. As a benchmark background estimate, we augment both spectra with the diffuse gamma-ray background from Table 3 of Ref.~\cite{Fermi-LAT:2014ryh}. We then perform a $\chi^2$ analysis to determine whether the two spectra are statistically distinguishable. This range and binning were chosen as a proof of concept. For particular models, the energy window can be tuned, while staying above $\sim 1$ GeV, to concentrate on bins with the largest significance between the BSM and SM models.

\Fig{fig:telescope_spec} presents an approximation for the minimum telescope properties, $T_\mathrm{obs}A_\mathrm{eff}$, needed for discovery of the model classes considered here, where $T_\mathrm{obs}$ is the observation time and $A_\mathrm{eff}$ the effective area. The models are the SM plus $N$ additional scalar particles of mass $m_\mathrm{dark}$ that only gravitationally interact with the SM.

As shown in \Fig{fig:telescope_spec}, sensitivity to heavier particles is lost because deviations from the SM signal become increasingly suppressed beneath the background as the flux drops. For $10^2~\mathrm{MeV} < m_\textrm{dark} < 1~\mathrm{GeV}$, sensitivity becomes weaker as spectral alterations occur at $E_\gamma \lesssim 1$ GeV, before the tail region can be conclusively identified. This leads to an overall suppression of the tail in the observation window compared with other models. Because the photon suppression grows with $N$, the order of model sensitivity reverses.

The low-mass region requires more care. Heavy particles ($m_\mathrm{dark}\gg T_\mathrm{BH}(\mevp)$) become available only late in the evaporation, when $T_\mathrm{BH}(M)\gtrsim m_\mathrm{dark}$, and suppress the tail by increasing $\varepsilon(M)$ in \Eq{eq:density_comp}. Light particles ($m_\mathrm{dark}\lesssim T_\mathrm{BH}(\mevp)$), however, also enter $\varepsilon(M_i)$ of the heavier progenitor PBHs, with two consequences. First, $\mevp$ shifts to higher masses, so distributions normalized at the SM $\mevp$ no longer share a common tail amplitude, which is readily corrected by adjusting $f_\mathrm{PBH}$ or $A$. Second, the tail is enhanced rather than suppressed. The light states contribute to $\varepsilon$ at both the initial and final BH mass. However, because the SM contains few light dof, the relative increase is more pronounced in $\varepsilon(M_i)$, enhancing the tail in \Eq{eq:density_comp} and making these models easier to detect. This enhancement asymptotes once $m_\mathrm{dark}\ll T_\mathrm{BH}(\mevp)$, where the particles are effectively massless.

\Fig{fig:telescope_spec} restricts itself to a single set of search criteria. However, as stated earlier, due to model properties or experimental limitations, other criteria might be preferred. For example, if all parts of the above analysis are maintained, but the photon energy range is shifted from $1$--$10~\mathrm{GeV}$ to $0.1$--$1~\mathrm{TeV}$, the telescopic requirements for the $N=10^2$ model from \Fig{fig:telescope_spec} are $10^{5}~{\rm m^2~yr}$ for light masses and $2\times10^{6}~{\rm m^2~yr}$ for heavy masses. See the {\it Supplemental Material} for a figure similar to \Fig{fig:telescope_spec} with this analysis specification.

\section{Discussion}

We have shown that the low-mass tail of the PBH mass function provides a robust probe of new degrees of freedom. If the PBH gamma-ray spectrum lies near current bounds, future gamma-ray experiments with square-meter-scale effective areas and 10-year observation time can probe $\mathcal{O}(10$--$10^3)$ new dof over a broad range of masses, with little dependence on the initial PBH mass spectrum. This effective area in the $1$--$10~{\rm m}^2$ range corresponds to the design goals of proposed missions such as APT~\cite{Buckley2019Advanced,APT:2021lhj,Alnussirat:2021tlo} and VLAST~\cite{Zhang:2024jha}. Additionally, for a photon search range of $0.1$--$1~\mathrm{TeV}$ over the same 10-year period, the necessary effective area is $10^4$--$10^7~{\rm m}^2$, comparable to instruments like the CTA observatory~\cite{Lopez-Oramas:2025vld}. However, for the results shown here, reaching this CTA sensitivity would require the evaporation tail's normalization to be determined by a separate instrument.

The distinctive feature of this probe is its insensitivity to how the new particles are realized. Because Hawking radiation is purely gravitational, the tail responds only to the number of light dof and their mass thresholds, regardless of their couplings to the visible sector. The signal is therefore present even for completely secluded states, and it does not depend on whether these states were populated in the early Universe. This sets the method apart from probes such as the effective number of relativistic species $N_\mathrm{eff}$ or collider and direct-detection searches, all of which require the new states to couple to or be produced alongside the visible sector. 

A large number of dof ($N$) may remain consistent with cosmological bounds because PBH evaporation can probe secluded particles that need not have been populated during reheating. For example, neutral naturalness scenarios such as the Mirror-Twin-Higgs~\cite{Chacko:2005pe,Chacko:2005un} and $N$naturalness models~\cite{Arkani-Hamed:2016rle} introduce $\mathcal{O}(10$--$10^4)$ dark-sector degrees of freedom to address the Higgs little-hierarchy problem. These sectors are only weakly populated due to asymmetric reheating and can satisfy existing cosmological constraints~\cite{Chacko:2018vss,Bansal:2021dfh,Zu:2023rmc,Bansal:2024afn}. Many UV theories also feature multiple shift symmetries that give rise to a large number of light ALPs, with $N \sim 10\text{--}100$. In the absence of sizable couplings to the Standard Model, these ALPs can remain underpopulated and evade cosmological bounds~\cite{Dessert:2025yvk}.

Our sensitivity estimates rest on a proof-of-concept $\chi^2$ analysis and a specific baseline PBH distribution, and a dedicated analysis tailored to a given mission would sharpen them. Even so, observing a modified Hawking radiation tail would provide a robust probe of large numbers of hidden degrees of freedom that are otherwise difficult to access. As next-generation gamma-ray missions approach the required sensitivity, the PBH evaporation tail may become one of the few practical ways to uncover such large particle multiplicities.

\section*{Acknowledgements}
We are grateful to Manuel Buen-Abad for valuable contributions in the early stages of this work. We also appreciate Yuber Perez-Gonzalez and Jessica Turner for providing the data from their work. The work of KA was supported by NSF grant PHY-2210361 and by the Maryland Center for Fundamental Physics. The work of BD is supported by the U.S. DOE Grant DE-SC0010813. TX acknowledges support from the HKUST Jockey Club Institute for Advanced Study. The research activity of TX is supported in part by the U.S. National Science Foundation under award PHY-2412671. YT is supported by the NSF Grant Number PHY2412701. YT would also like to thank the Aspen Center for Physics (supported by NSF grant PHY-2210452) and the Munich Institute for Astro-, Particle and BioPhysics (MIAPbP), which is
funded by the Deutsche Forschungsgemeinschaft (DFG, German Research Foundation) under
Germany’s Excellence Strategy – EXC-2094 – 390783311.

\bibliographystyle{utphys}
\bibliography{reference.bib}

@article{Boluna:2023jlo,
    author = "Boluna, Xavier and Profumo, Stefano and Bl\'e, Juliette and Hennings, Dana",
    title = "{Searching for Exploding black holes}",
    eprint = "2307.06467",
    archivePrefix = "arXiv",
    primaryClass = "astro-ph.HE",
    doi = "10.1088/1475-7516/2024/04/024",
    journal = "JCAP",
    volume = "04",
    pages = "024",
    year = "2024"
}

@article{Foot:2014mia,
    author = "Foot, R.",
    title = "{Mirror dark matter: Cosmology, galaxy structure and direct detection}",
    eprint = "1401.3965",
    archivePrefix = "arXiv",
    primaryClass = "astro-ph.CO",
    doi = "10.1142/S0217751X14300130",
    journal = "Int. J. Mod. Phys. A",
    volume = "29",
    pages = "1430013",
    year = "2014"
}

@article{Chacko:2005un,
    author = "Chacko, Z. and Goh, Hock-Seng and Harnik, Roni",
    title = "{A Twin Higgs model from left-right symmetry}",
    eprint = "hep-ph/0512088",
    archivePrefix = "arXiv",
    reportNumber = "SLAC-PUB-11595",
    doi = "10.1088/1126-6708/2006/01/108",
    journal = "JHEP",
    volume = "01",
    pages = "108",
    year = "2006"
}

@article{Chacko:2005pe,
    author = "Chacko, Z. and Goh, Hock-Seng and Harnik, Roni",
    title = "{The Twin Higgs: Natural electroweak breaking from mirror symmetry}",
    eprint = "hep-ph/0506256",
    archivePrefix = "arXiv",
    doi = "10.1103/PhysRevLett.96.231802",
    journal = "Phys. Rev. Lett.",
    volume = "96",
    pages = "231802",
    year = "2006"
}

@article{Lopez-Oramas:2025vld,
    author = "L{\'o}pez-Oramas, Alicia",
    collaboration = "CTAO",
    title = "{CTAO status and perspective}",
    doi = "10.1051/epjconf/202531901002",
    journal = "EPJ Web Conf.",
    volume = "319",
    pages = "01002",
    year = "2025"
}

@article{Alnussirat:2021tlo,
    author = "Alnussirat, Samer and others",
    title = "{The Advanced Particle-astrophysics Telescope: Simulation of the Instrument Performance for Gamma-Ray Detection}",
    doi = "10.22323/1.395.0590",
    journal = "PoS",
    volume = "ICRC2021",
    pages = "590",
    year = "2021"
}

@article{Buckley2019Advanced,
	author = {Buckley, James and Bergstrom, Lars and Binns, Bob and Buhler, Jeremy and Chen, Wenlei and Cherry, Michael and Funk, Stefan and Hooper, Dan and Mitchell, John and Nolfo, Georgia De and Nussirat, Samer Al and Profumo, Stefano and Rauch, Brian and Stern, Daniel and Varner, Garry and Wakely, Scott and Zink, Adrian},
	journal = {Bulletin of the AAS},
	number = {7},
	year = {2019},
	month = {sep 30},
	note = {https://baas.aas.org/pub/2020n7i078},
	publisher = {},
	title = {The {Advanced} {Particle}-astrophysics {Telescope} ({APT})},
	volume = {51},
}

@article{APT:2021lhj,
    author = "Buckley, James H. and others",
    collaboration = "APT",
    title = "{The Advanced Particle-astrophysics Telescope (APT) Project Status}",
    doi = "10.22323/1.395.0655",
    journal = "PoS",
    volume = "ICRC2021",
    pages = "655",
    year = "2021"
}

@article{Zhang:2024jha,
    author = "Zhang, Yanshuo",
    collaboration = "VLAST",
    title = "{Previous studies about BGO calorimeter in Very Large Area gamma ray Space Telescope}",
    doi = "10.22323/1.444.0897",
    journal = "PoS",
    volume = "ICRC2023",
    pages = "897",
    year = "2024"
}

@article{Coogan:2020tuf,
    author = "Coogan, Adam and Morrison, Logan and Profumo, Stefano",
    title = "{Direct Detection of Hawking Radiation from Asteroid-Mass Primordial Black Holes}",
    eprint = "2010.04797",
    archivePrefix = "arXiv",
    primaryClass = "astro-ph.CO",
    doi = "10.1103/PhysRevLett.126.171101",
    journal = "Phys. Rev. Lett.",
    volume = "126",
    number = "17",
    pages = "171101",
    year = "2021"
}

@article{Chacko:2018vss,
    author = "Chacko, Zackaria and Curtin, David and Geller, Michael and Tsai, Yuhsin",
    title = "{Cosmological Signatures of a Mirror Twin Higgs}",
    eprint = "1803.03263",
    archivePrefix = "arXiv",
    primaryClass = "hep-ph",
    reportNumber = "FERMILAB-PUB-18-075-T",
    doi = "10.1007/JHEP09(2018)163",
    journal = "JHEP",
    volume = "09",
    pages = "163",
    year = "2018"
}

@article{Zu:2023rmc,
    author = "Zu, Lei and Zhang, Chi and Chen, Hou-Zun and Wang, Wei and Tsai, Yue-Lin Sming and Tsai, Yuhsin and Luo, Wentao and Fan, Yi-Zhong",
    title = "{Exploring mirror twin Higgs cosmology with present and future weak lensing surveys}",
    eprint = "2304.06308",
    archivePrefix = "arXiv",
    primaryClass = "astro-ph.CO",
    doi = "10.1088/1475-7516/2023/08/023",
    journal = "JCAP",
    volume = "08",
    pages = "023",
    year = "2023"
}

@article{Bansal:2024afn,
    author = "Bansal, Saurabh and Ghosh, Subhajit and Low, Matthew and Tsai, Yuhsin",
    title = "{A cosmological case study of a tower of warm dark matter states: Nnaturalness}",
    eprint = "2410.19224",
    archivePrefix = "arXiv",
    primaryClass = "astro-ph.CO",
    reportNumber = "UTWI-31-2024",
    doi = "10.1088/1475-7516/2025/11/012",
    journal = "JCAP",
    volume = "11",
    pages = "012",
    year = "2025"
}

@article{Bansal:2021dfh,
    author = "Bansal, Saurabh and Kim, Jeong Han and Kolda, Christopher and Low, Matthew and Tsai, Yuhsin",
    title = "{Mirror twin Higgs cosmology: constraints and a possible resolution to the H$_{0}$ and S$_{8}$ tensions}",
    eprint = "2110.04317",
    archivePrefix = "arXiv",
    primaryClass = "hep-ph",
    reportNumber = "FERMILAB-PUB-21-371-T",
    doi = "10.1007/JHEP05(2022)050",
    journal = "JHEP",
    volume = "05",
    pages = "050",
    year = "2022"
}

@article{Dessert:2025yvk,
    author = "Dessert, Christopher and Kumar, Soubhik and Ruderman, Joshua T.",
    title = "{Freezing-in the Axiverse}",
    eprint = "2511.09631",
    archivePrefix = "arXiv",
    primaryClass = "hep-ph",
    month = "11",
    year = "2025"
}

@article{Baker:2025ffi,
    author = "Baker, Michael J. and Iguaz Juan, Joaquim and Symons, Aidan and Thamm, Andrea",
    title = "{Probing Dark Sectors with Exploding Black Holes: Gamma Rays}",
    eprint = "2512.19603",
    archivePrefix = "arXiv",
    primaryClass = "hep-ph",
    month = "12",
    year = "2025"
}

@article{Arbey:2019mbc,
    author = "Arbey, Alexandre and Auffinger, J\'er\'emy",
    title = "{BlackHawk: A public code for calculating the Hawking evaporation spectra of any black hole distribution}",
    eprint = "1905.04268",
    archivePrefix = "arXiv",
    primaryClass = "gr-qc",
    reportNumber = "CERN-TH-2019-067",
    doi = "10.1140/epjc/s10052-019-7161-1",
    journal = "Eur. Phys. J. C",
    volume = "79",
    number = "8",
    pages = "693",
    year = "2019"
}

@article{Arbey:2021mbl,
    author = "Arbey, Alexandre and Auffinger, J\'er\'emy",
    title = "{Physics Beyond the Standard Model with BlackHawk v2.0}",
    eprint = "2108.02737",
    archivePrefix = "arXiv",
    primaryClass = "gr-qc",
    reportNumber = "CERN-TH-2021-117",
    doi = "10.1140/epjc/s10052-021-09702-8",
    journal = "Eur. Phys. J. C",
    volume = "81",
    pages = "910",
    year = "2021"
}

@article{Baker:2021btk,
    author = "Baker, Michael J. and Thamm, Andrea",
    title = "{Probing the particle spectrum of nature with evaporating black holes}",
    eprint = "2105.10506",
    archivePrefix = "arXiv",
    primaryClass = "hep-ph",
    doi = "10.21468/SciPostPhys.12.5.150",
    journal = "SciPost Phys.",
    volume = "12",
    number = "5",
    pages = "150",
    year = "2022"
}

@article{Baker:2022rkn,
    author = "Baker, Michael J. and Thamm, Andrea",
    title = "{Black hole evaporation beyond the Standard Model of particle physics}",
    eprint = "2210.02805",
    archivePrefix = "arXiv",
    primaryClass = "hep-ph",
    doi = "10.1007/JHEP01(2023)063",
    journal = "JHEP",
    volume = "01",
    pages = "063",
    year = "2023"
}

@article{Cheek:2022mmy,
    author = "Cheek, Andrew and Heurtier, Lucien and Perez-Gonzalez, Yuber F. and Turner, Jessica",
    title = "{Evaporation of primordial black holes in the early Universe: Mass and spin distributions}",
    eprint = "2212.03878",
    archivePrefix = "arXiv",
    primaryClass = "hep-ph",
    reportNumber = "IPPP/22/79",
    doi = "10.1103/PhysRevD.108.015005",
    journal = "Phys. Rev. D",
    volume = "108",
    number = "1",
    pages = "015005",
    year = "2023"
}

@article{Page:1976ki,
    author = "Page, Don N.",
    title = "{Particle Emission Rates from a Black Hole. 2. Massless Particles from a Rotating Hole}",
    doi = "10.1103/PhysRevD.14.3260",
    journal = "Phys. Rev. D",
    volume = "14",
    pages = "3260--3273",
    year = "1976"
}

@article{Navarro:1996gj,
    author = "Navarro, Julio F. and Frenk, Carlos S. and White, Simon D. M.",
    title = "{A Universal density profile from hierarchical clustering}",
    eprint = "astro-ph/9611107",
    archivePrefix = "arXiv",
    doi = "10.1086/304888",
    journal = "Astrophys. J.",
    volume = "490",
    pages = "493--508",
    year = "1997"
}

@article{Carr:1974nx,
    author = "Carr, Bernard J. and Hawking, S. W.",
    title = "{Black holes in the early Universe}",
    doi = "10.1093/mnras/168.2.399",
    journal = "Mon. Not. Roy. Astron. Soc.",
    volume = "168",
    pages = "399--415",
    year = "1974"
}

@article{Hawking:1975vcx,
    author = "Hawking, S. W.",
    editor = "Gibbons, G. W. and Hawking, S. W.",
    title = "{Particle Creation by Black Holes}",
    doi = "10.1007/BF02345020",
    journal = "Commun. Math. Phys.",
    volume = "43",
    pages = "199--220",
    year = "1975",
    note = "[Erratum: Commun.Math.Phys. 46, 206 (1976)]"
}

@article{Mosbech:2022lfg,
    author = "Mosbech, Markus R. and Picker, Zachary S. C.",
    title = "{Effects of Hawking evaporation on PBH distributions}",
    eprint = "2203.05743",
    archivePrefix = "arXiv",
    primaryClass = "astro-ph.HE",
    doi = "10.21468/SciPostPhys.13.4.100",
    journal = "SciPost Phys.",
    volume = "13",
    number = "4",
    pages = "100",
    year = "2022"
}

@article{Klipfel:2025bvh,
    author = "Klipfel, Alexandra P. and Fisher, Peter and Kaiser, David I.",
    title = "{Hawking radiation signatures from primordial black holes transiting the inner Solar System: Prospects for detection}",
    eprint = "2506.14041",
    archivePrefix = "arXiv",
    primaryClass = "astro-ph.CO",
    reportNumber = "Preprint MIT-CTP/5878",
    doi = "10.1103/9jyp-24sw",
    journal = "Phys. Rev. D",
    volume = "112",
    number = "10",
    pages = "103007",
    year = "2025"
}

@article{Carr:2026hot,
    author = {Carr, Bernard and Iovino, Antonio J. and Perna, Gabriele and Vaskonen, Ville and Veerm{\"a}e, Hardi},
    title = "{Primordial black holes: constraints, potential evidence and prospects}",
    eprint = "2601.06024",
    archivePrefix = "arXiv",
    primaryClass = "astro-ph.CO",
    doi = "10.1007/s40766-026-00080-z",
    month = "1",
    year = "2026"
}

@article{Arkani-Hamed:2016rle,
    author = "Arkani-Hamed, Nima and Cohen, Timothy and D'Agnolo, Raffaele Tito and Hook, Anson and Kim, Hyung Do and Pinner, David",
    title = "{Solving the Hierarchy Problem at Reheating with a Large Number of Degrees of Freedom}",
    eprint = "1607.06821",
    archivePrefix = "arXiv",
    primaryClass = "hep-ph",
    doi = "10.1103/PhysRevLett.117.251801",
    journal = "Phys. Rev. Lett.",
    volume = "117",
    number = "25",
    pages = "251801",
    year = "2016"
}

@article{deSalas:2019pee,
    author = "de Salas, P. F. and Malhan, K. and Freese, K. and Hattori, K. and Valluri, M.",
    title = "{On the estimation of the Local Dark Matter Density using the rotation curve of the Milky Way}",
    eprint = "1906.06133",
    archivePrefix = "arXiv",
    primaryClass = "astro-ph.GA",
    doi = "10.1088/1475-7516/2019/10/037",
    journal = "JCAP",
    volume = "10",
    pages = "037",
    year = "2019"
}

@article{Arvanitaki:2009fg,
    author = "Arvanitaki, Asimina and Dimopoulos, Savas and Dubovsky, Sergei and Kaloper, Nemanja and March-Russell, John",
    title = "{String Axiverse}",
    eprint = "0905.4720",
    archivePrefix = "arXiv",
    primaryClass = "hep-th",
    doi = "10.1103/PhysRevD.81.123530",
    journal = "Phys. Rev. D",
    volume = "81",
    pages = "123530",
    year = "2010"
}

@article{Fermi-LAT:2014ryh,
    author = "Ackermann, M. and others",
    collaboration = "Fermi-LAT",
    title = "{The spectrum of isotropic diffuse gamma-ray emission between 100 MeV and 820 GeV}",
    eprint = "1410.3696",
    archivePrefix = "arXiv",
    primaryClass = "astro-ph.HE",
    doi = "10.1088/0004-637X/799/1/86",
    journal = "Astrophys. J.",
    volume = "799",
    pages = "86",
    year = "2015"
}

@article{Fermi-LAT:2009ihh,
    author = "Atwood, W. B. and others",
    collaboration = "Fermi-LAT",
    title = "{The Large Area Telescope on the Fermi Gamma-ray Space Telescope Mission}",
    eprint = "0902.1089",
    archivePrefix = "arXiv",
    primaryClass = "astro-ph.IM",
    reportNumber = "SLAC-PUB-13620",
    doi = "10.1088/0004-637X/697/2/1071",
    journal = "Astrophys. J.",
    volume = "697",
    pages = "1071--1102",
    year = "2009"
}

@book{Raffelt:1996wa,
    author = "Raffelt, G. G.",
    title = "{Stars as laboratories for fundamental} {physics: The astrophysics of neutrinos, axions, and} {other weakly interacting particles}",
    isbn = "978-0-226-70272-8",
    month = "5",
    year = "1996"
}

@article{Raffelt:1990yz,
    author = "Raffelt, Georg G.",
    title = "{Astrophysical methods to constrain axions and other novel particle phenomena}",
    reportNumber = "MPI-PAE-PTH-29-90",
    doi = "10.1016/0370-1573(90)90054-6",
    journal = "Phys. Rept.",
    volume = "198",
    pages = "1--113",
    year = "1990"
}

@article{Caputo:2022xpx,
    author = "Caputo, Regina and others",
    title = "{All-sky Medium Energy Gamma-ray Observatory eXplorer mission concept}",
    eprint = "2208.04990",
    archivePrefix = "arXiv",
    primaryClass = "astro-ph.IM",
    doi = "10.1117/1.JATIS.8.4.044003",
    journal = "J. Astron. Telesc. Instrum. Syst.",
    volume = "8",
    number = "4",
    pages = "044003",
    year = "2022"
}

@article{He:2026neq,
    author = "He, Xin-Chen and Ma, Xiao-Han and Sasaki, Misao and Takhistov, Volodymyr",
    title = "{Universal Suppression of Gravitational Waves from Black Hole Evaporation Dynamics}",
    eprint = "2606.09804",
    archivePrefix = "arXiv",
    primaryClass = "astro-ph.CO",
    reportNumber = "YITP-26-58, KEK-QUP-2026-0010, KEK-TH-2844",
    month = "6",
    year = "2026"
}

@article{Page:1976wx,
    author = "Page, Don N. and Hawking, S. W.",
    title = "{Gamma rays from primordial black holes}",
    doi = "10.1086/154350",
    journal = "Astrophys. J.",
    volume = "206",
    pages = "1--7",
    year = "1976"
}

@article{Halzen:1991uw,
    author = "Halzen, F. and Zas, E. and MacGibbon, J. H. and Weekes, T. C.",
    title = "{Gamma-rays and energetic particles from primordial black holes}",
    doi = "10.1038/353807a0",
    journal = "Nature",
    volume = "353",
    pages = "807--815",
    year = "1991"
}

\widetext
\begin{center}
\textbf{\large Supplemental Material}
\end{center}
\setcounter{equation}{0}
\setcounter{figure}{0}
\setcounter{table}{0}
\setcounter{page}{1}
\makeatletter
\renewcommand{\theequation}{S\arabic{equation}}
\renewcommand{\thefigure}{S\arabic{figure}}
\renewcommand{\bibnumfmt}[1]{[#1]}
\renewcommand{\citenumfont}[1]{#1}

\section{The Evolution of PBH Mass Functions}

In order to discuss BH distributions across different epochs, it is important to understand how the BH distribution function evolves. To begin, we first note that the number of BHs ($N$) in a comoving volume $V$ and mass window $M \in [M_1,M_2]$ is
\begin{equation}
    N = V\int_{M_1}^{M_2} \hat{n} \; \mathrm{d}M \,,
\end{equation}
where $\hat{n} \equiv a^3 \mathrm{d}n/\mathrm{d}M$ is the comoving BH number density per mass with the scale factor $a$. Note that $\hat{n}$ is equivalent to $\hat{n}_{\BH}$ from the main text; the subscripts have been removed for notation simplicity. As the BHs evaporate, it is obvious but important to note that the mass of an individual BH changes; however, the total number of BHs is conserved as long as their new mass is nonzero. Let $M_0(\Delta t,M_i)$ represent the mass of a black hole at time $t_0$ that was of mass $M_i$ at time $t=t_i$ where $\Delta t \equiv t_0 - t_i$ is the evolution time. As previously stated, as the BHs evaporate, the total number of BHs during different time periods is constant and results in the relationship 
\begin{equation}
    N = V\int_{M_{i,1}}^{M_{i,2}} \hat{n}_i \; \mathrm{d}M_i = V\int_{M_{0,1}}^{M_{0,2}} \hat{n}_0 \; \mathrm{d}M_0 \,,
\end{equation}
where the indices (``0'' and ``$i$'') refer to quantities derived at their respective time periods ($t_0$ and $t_i$) and the ``1'' and ``2'' indices correspond to the edges of the mass window. For an infinitesimally small mass window, the relationship between the initial number distribution and a later one is found to be
\begin{equation}
    \hat{n}_0 = \frac{\Delta M_i}{\Delta M_0} \hat{n}_i = \frac{\mathrm{d} M_i}{\mathrm{d} M_0} \hat{n}_i
    \label{eq:density_comp_temp} \,,
\end{equation}
where $\Delta M$ is the size of the mass window, with indices (``$i$'' and ``0'') again noting the time period to which the quantity refers. The indices also indicate that the quantity is calculated at the BH mass corresponding to that particular era ($M_i$ and $M_0$ respectively).

The evolution time, $\Delta t$, is common to all black holes present today, so the initial and final masses are related by
\begin{equation}
    \Delta t = \int_{M_i}^{M_0} \frac{\mathrm{d}M}{\dot{M}(M)} \, .\label{eq:t_of_M}
\end{equation}
Since $\Delta t$ is the same for all present-day black holes and $\dot{M} = \dot{M}(M)$ has no explicit time dependence, differentiating both sides with respect to $M_0$ and using $\mathrm{d}\Delta t/\mathrm{d}M_0 = 0$ yields
\begin{equation}
    0 = \frac{1}{\dot{M}_0} - \frac{1}{\dot{M}_i}\frac{\mathrm{d}M_i}{\mathrm{d}M_0} \, ,
\end{equation}
which rearranges to
\begin{equation}
    \frac{\mathrm{d} M_i}{\mathrm{d} M_0} = \frac{\dot{M}_i}{\dot{M}_0} \, ,
    \label{eq:bin_ratio}
\end{equation}
where $\dot{M}(M_a)=\dot{M}_a$ and, for use below, $\varepsilon(M_a) = \varepsilon_a$ are introduced for notation simplicity. Substituting \Eq{eq:bin_ratio} into \Eq{eq:density_comp_temp} yields the simplified number density relationship
\begin{equation}
   \hat{n}_0 = \frac{\dot{M}_i}{\dot{M}_0} \hat{n}_i = \frac{\varepsilon_i}{\varepsilon_0} \frac{M_0^2}{M_i^2} \hat{n}_i \, .
   \label{eq:density_comp2}
\end{equation}

Evolving a mass distribution thus requires simply mapping the initial BH masses to their future values (evaluating $M_0(\Delta t,M_i)$) and adjusting the initial mass distribution by the ratio of the two evaporation rates. In practice however, it is more convenient to define the reverse function $M_i(\Delta t,M_0)$, to find the initial mass $M_i$ given the current mass $M_0$ and $\Delta t$. If multiple $\Delta t$ are required, a simple approach to solving $M_i(\Delta t,M_0)$ is to numerically evaluate \Eq{eq:t_of_M} with $M_i=M$ and $M_0=0$ at various $M$ to find the evaporation time $t_\mathrm{evp}(M)$ and the inverse function $\mevp (t)$, the BH mass that will evaporate in time $t$. Since $t_\mathrm{evp}(M_i)=t_\mathrm{evp}(M_0)+\Delta t$,
\begin{equation}
   M_i(\Delta t,M_0) = \mevp (t_\mathrm{evp}(M_0)+ \Delta t) \, .
   \label{eq:initial_mass_equation}
\end{equation}

For masses deep in the evaporation tail, $M_i\approx\mevp$, and \Eq{eq:density_comp2} reduces to
\begin{equation}
    \hat{n}_0(M_0 \ll M_{\rm evp})\approx \frac{M_0^2}{\varepsilon_0} \frac{\varepsilon_\mathrm{evp}}{\mevp^2} \hat{n}_\mathrm{evp} =\left({\rm overall\,\,amplitude}\right)\left[\frac{M_0^2}{\varepsilon_0} \right] \, ,
    \label{eq:density_comp_deep_tail}
\end{equation}
the universal evaporation tail whose only dependence on the initial PBH mass distribution is an overall amplitude.

For most of a BH's lifetime, $\varepsilon = \varepsilon_i$. In this limit, $M_i^3\approx M_0^3 + 3 \varepsilon_i \mPl^4 \Delta t$, and the number density relationship becomes
\begin{equation}
    \left.\hat{n}_0 \middle/ \hat{n}_i \right. = \frac{\varepsilon_i}{\varepsilon_0} \frac{M_0^2}{(M_0^3 + 3 \varepsilon_i \mPl^4 \Delta t)^{2/3}} \, ,
    \label{eq:density_comp_limit}
\end{equation}
similar to results in other works~\cite{Page:1976wx, Halzen:1991uw, Mosbech:2022lfg, Klipfel:2025bvh, Boluna:2023jlo}; however, because these works approximate $M_i(\Delta t, M_0)$ before arriving at \Eq{eq:density_comp2}, the ratio in $\varepsilon$ was omitted. The $\varepsilon$ term is crucial for modulating the distribution as new dof become accessible.

While \Eq{eq:density_comp_limit} has limited use when compared with \Eq{eq:density_comp2} and \Eq{eq:density_comp_deep_tail} due to its continued dependence on $M_i$ through $\varepsilon_i$, it can be illustrative when comparing the limiting behavior of models with different $\varepsilon(M)$, especially those where the additional dof modify $\mevp$. Taking the ratio of \Eq{eq:density_comp_limit} for two different $\varepsilon(M)$ models denoted by the numerical index at the same $M_0$ and $\Delta t$, the relative number ratio is
\begin{equation}
    \frac{\left[\hat{n}_0 \middle/ \hat{n}_i \right]_2}{\left[\hat{n}_0 \middle/ \hat{n}_i \right]_1} = \frac{\varepsilon_{i,2}}{\varepsilon_{i,1}} \frac{\varepsilon_{0,1}}{\varepsilon_{0,2}} \left(\frac{M_0^3 + 3 \varepsilon_{i,1} \mPl^4 \Delta t}{M_0^3 + 3 \varepsilon_{i,2} \mPl^4 \Delta t}\right)^{2/3}. \label{eq:density_comp_limit_ratio}
\end{equation}
For $\Delta t\gg M_0^3/3 \varepsilon_{i,a} \mPl^4$ (the deep tail limit with constant $\mevp$), this can be further simplified to
\begin{equation}
    \frac{\left[\hat{n}_0 \middle/ \hat{n}_i \right]_2}{\left[\hat{n}_0 \middle/ \hat{n}_i \right]_1} = \frac{\varepsilon_{0,1}}{\varepsilon_{0,2}} \left(\frac{\varepsilon_{i,2}}{\varepsilon_{i,1}}\right)^{1/3} .\label{eq:density_comp_limit_ratio_simp}
\end{equation}

As expected, adding heavy dof (additional particles that only become active at the end of a BH's life, increasing $\varepsilon_0$) results in a drop in the relative ratio between the two models. However, adding light dof (always massless when compared with the initial temperature of the BH) may have the opposite effect and cause an increase in the relative ratio. This increase is due to contributions to the evaporation rate from these early extra dof outweighing the suppression that will also occur from increases to $M_i$. Altogether, these effects result in an increase in the initial mass window. This enhancement leads to the curious BH distribution behavior where models with larger evaporation rates may develop larger relative tails.

\section{Variations in Telescope Sensitivity with Energy Range}
\begin{figure}[ht!]
\includegraphics[width=0.9\columnwidth]{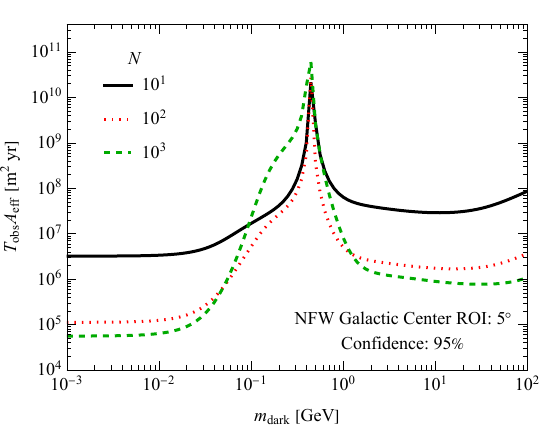}
\caption{Minimum telescopic properties for utilizing the universal PBH evaporation tail to discover dark BSM models using $0.1$--$1~\mathrm{TeV}$ photons.}
\label{fig:telescope_spec2}
\end{figure}
Similar to \Fig{fig:telescope_spec}, \Fig{fig:telescope_spec2} is an estimate of the minimum observation time ($T_\mathrm{obs}$) times effective area ($A_\mathrm{eff}$) to discover dark BSM models using the tail of a PBH distribution. The analysis incorporates all of the same restrictions on the PBH mass spectrum, background, and comparison model as the analysis performed for \Fig{fig:telescope_spec}. These are the experimental constraint on the mass distribution amplitude $A$ ($E_\gamma^2 {\rm d}\Phi_\gamma / {\rm d}E_\gamma=2.5\times 10^{-5}~{\rm MeV~cm^{-2}~s^{-1}}$ at 100 MeV), the SM comparison rescaling energy (1 GeV), and the spectral bin resolution (10 bins per decade). However, the analysis is performed over a different photon energy window $0.1$--$1~\mathrm{TeV}$. All of the same trends observed in \Fig{fig:telescope_spec} are present here. The main difference is a decrease in sensitivity due to the decrease in photon flux at these larger energies. The sensitivities and energy range shown here are comparable to those of CTA~\cite{Lopez-Oramas:2025vld}.

\end{document}